\documentclass[fleqn,twoside]{article}
\usepackage[headings]{espcrc2}

\readRCS
$Id: espcrc2.tex,v 1.2 2004/02/24 11:22:11 spepping Exp $
\usepackage{graphicx}
\usepackage[figuresright]{rotating}

\newcommand{\AmS}{{\protect\the\textfont2
  A\kern-.1667em\lower.5ex\hbox{M}\kern-.125emS}}

\title{Pentaquark Searches in Electron-Positron Annihilations and\\ Two-Photon Collisions at LEP}

\author{S.R. Armstrong\address[MCSD]{Brookhaven National Laboratory, Department of Physics,\\ 
        Upton, New York 11973-5000, United States of America}}
       
\runtitle{Pentaquark Searches at LEP}
\runauthor{S.R. Armstrong}

\begin{document}

\begin{abstract}
Evidence for the production of exotic hadron states composed of five
quarks (pentaquarks) has been searched for in data collected by the
ALEPH, DELPHI, and L3 experiments at LEP.  No significant signal is
observed.  Several 95\% C.L. upper limits are set on the production
rates of such particles.
\vspace{1pc}
\end{abstract}

\maketitle

\section{Introduction}
The past 18 months have seen extraordinary developments in the field of
hadron spectroscopy.  Several experiments have reported strong
experimental evidence of narrow exotic baryon resonances with quantum
numbers indicative of the existence of bound states beyond the
canonical constituent-quark model view with two or three (anti-)quark
bound states.  Permitted within the framework of Quantum Chromodynamics, these
states have been previously unobserved.

Numerous experimental excesses have been reported in
pK$^{0}_{\mathrm{S}}$ and nK$^+$ invariant mass spectra~\cite{obs}
consistent with a resonance of mass near 1535\,MeV/$c^2$ and width
less than 10\,MeV/$c^2$ carrying strangeness $S = +1$.\footnote{Here, and throughout this
paper, charge conjugation is assumed.}  Furthermore,
evidence of a narrow resonance in $\Xi^{-}\pi^{\pm}$ invariant mass
spectra has also been observed~\cite{cascadeobs} consistent with a
resonance of mass near 1860\,MeV/$c^2$ and width less than
18\,MeV/$c^2$.  Finally, the heavy flavor sector is not immune from
the appearance of these new states: an excess is observed in pD$^*-$
invariant mass spectra~\cite{h1obs} consistent with a resonance of mass near
3100\,MeV/$c^2$ and width less than 15\,MeV/$c^2$.

These experimental results have generated at least an
order-of-magnitude more papers concerning the
interpretation and phenomenology of these states~\cite{lanlpenta}.  A favored
interpretation for these new resonances is that of a {\it pentaquark}
bound state consisting of four quarks and an anti-quark ({\it e.g.},
udud$\overline{\mathrm{S}}$, dsds$\overline{\mathrm{u}}$, udud$\overline{\mathrm{c}}$); these
states are also described within and may enhance the validity of chiral
soliton models~\cite{csm}.

Searches for pentaquark states in e$^+$e$^-$ annihilations and $\gamma\gamma$ collisions
have been proposed~\cite{mynote}.  Results from these searches may be compared 
directly to the observation of known non-exotic states which decay into relevant
and similar final states ({\it e.g.}, $\Lambda(1520)\rightarrow\mathrm{pK}^{-}$, 
$\Xi(1530)^{0}\rightarrow\Xi^{-}\pi^{+}$, D$^{*+}$(2010)$\rightarrow$D$^{0}\pi^{+}$).
Negative results may help illuminate the as yet poorly understood pentaquark production
process.

\section{The Experiments and Data Sets}
From 1989 to 1995, LEP accelerated and collided counter-rotating beams
of electrons and positrons to energies of roughly 45\,GeV, allowing
collisions with center-of-mass energies at or near the Z peak.  This
first phase of LEP (LEP1) delivered a total of 200\,pb$^{-1}$ to each of
four large general-purpose detector experiments (ALEPH, DELPHI, L3,
and OPAL).  In these data, each experiment recorded more than 3
million hadronic Z decays yielding in excess of 5$\times 10^5$
reconstructed K$^{0}_{\mathrm{S}}$ mesons and 10$^4$ $\Lambda$
baryons.

The second phase of LEP (LEP2) involved increasing the center-of-mass
energies to nearly 210\,GeV in consecutive annual upgrades between
1995 and 2000.  Starting in November 1995, LEP attained center-of-mass
energies between 130 and 136\,GeV.  Subsequent years yielded energies
of 161 and 172\,GeV~(1996), 183\,GeV~(1997), 189\,GeV~(1998), between
192 and 196\,GeV~(1999), and between 200 and 209\,GeV~(2000).  An
integrated luminosity of about 500\,pb$^{-1}$ was delivered during LEP2
to each experiment.

Thus far, ALEPH, DELPHI, and L3 have analyzed various data samples
searching for evidence of pentaquark production.  These detectors are
described in detail elsewhere~\cite{alephdet,delphidet,l3det}.

\section{Searches for Pentaquark Production}
The event selection strategies across the three LEP experiments which
have conducted pentaquark searches are similar.  ALEPH~\cite{ALEPH} and DELPHI~\cite{DELPHI}
searched for evidence of pentaquark production from the fragmentation
of quarks from roughly four million hadronic Z decays
(e$^+$e$^{-}\rightarrow\mathrm{Z}\rightarrow\mathrm{q}\overline{\mathrm{q}}\rightarrow\Theta(1535)^{+}X$)
within data taken at or near center-of-mass energies of 91.2\,GeV.  L3~\cite{L3}
searched for evidence of pentaquark production from two-photon
interactions
(e$^+$e$^{-}\rightarrow\mathrm{e}^+\mathrm{e}^-\gamma\gamma\rightarrow\mathrm{e}^+\mathrm{e}^-\Theta^{+}X$)
in data taken between center-of-mass energies of 189 and 209\,GeV,
with a luminosity-weighted average value of 198\,GeV; the data sample
corresponds to 610\,pb$^{-1}$.

All of the experiments make quality cuts to select good events as well
as exploiting the relevant particle identification capabilities of
sub-detectors.  All of the experiments rely upon specific energy loss
by ionization (d$E$/d$x$) of charged particles traversing the fiducial
volume of various tracking chambers to discriminate between charged
particle species ($\pi^{\pm}$, K$^{\pm}$, p/$\overline{\mathrm{p}}$,
or e$^{\pm}$).  DELPHI complements such information with data from a
Barrel Ring Imaging Cherenkov Counter (BRICH).

Selection of K$^{0}_{\mathrm{S}}\rightarrow\pi^+\pi^-$ candidates
makes use of decay length criteria relative to the primary vertex of
oppositely-charged pairs of identified pions as well as cuts on the
quality of reconstructed secondary vertices.  Similarly, ALEPH selects
$\Lambda\rightarrow$p$\pi^-$ candidates by associating oppositely-charged
identified pion and proton tracks.  Furthermore, $\Lambda$ candidates
are paired with pion tracks which yield a displaced secondary vertex to yield
a sample of $\Xi^-$ candidates.  Finally, samples of D$^0$, D$^+$, and D$^{*+}$ mesons 
were obtained using techniques described in detail in Ref.~\cite{dmeson} with the
addition of harder momenta criteria ($p(\mathrm{D}^0)>7$\,GeV/$c$ and $p(\mathrm{D}^+)>14$\,GeV/$c$).

\subsection{Searches in pK channels}
\begin{figure}[tbp]
\includegraphics[width=7.5cm]{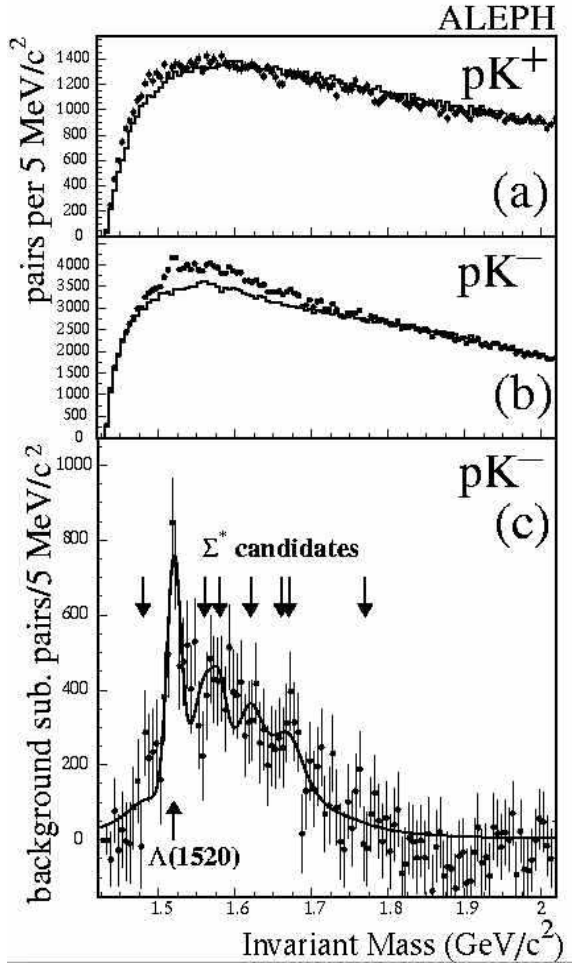}
\caption{\label{fig:alephpk}ALEPH invariant mass spectra for (a)
doubly-charged pK$^+$ combinations, (b) neutral pK$^-$ combinations,
(c) neutral pK$^-$ combinations after background subtractions with
locations of known $\Lambda(1520)$ and $\Sigma^*$ baryon resonances
indicated.  Data are denoted by the points with error bars.  In (a)
and (b), the solid line denotes Monte Carlo simulation including known
decuplet and octet baryons; in (c), the solid line represents a fit
which includes the known $\Lambda(1520)$ and $\Sigma^*$ resonances.}
\end{figure}
\begin{figure}[tbp]
\includegraphics[width=7.5cm]{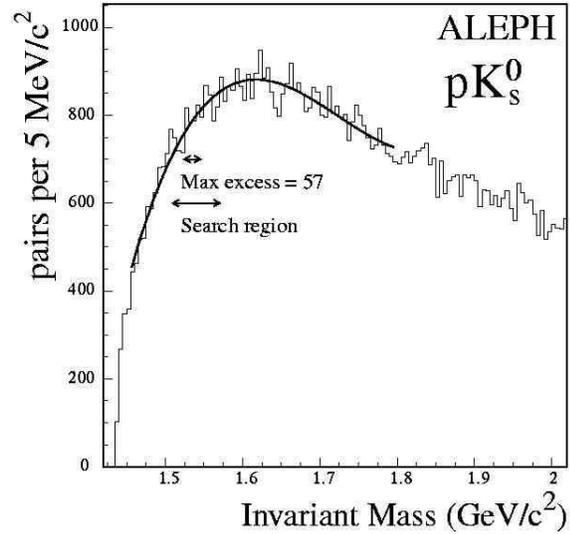}
\caption{ALEPH invariant mass spectrum for p$K^{0}_{\mathrm{S}}$ combinations.  A search
region indicated by the long horizontal area was examined for evidence of the $\Theta(1535)^+$.}
\label{fig:alephpk0s}
\end{figure}

As a benchmark and cross-check, ALEPH performed a search for
doubly-charged (pK$^+$) and neutral (pK$^-$) combinations.  No
resonant structure is observed in the invariant mass distribution of
doubly-charged pK combinations, shown in Figure~\ref{fig:alephpk}(a).
A smooth deviation of the data from Monte Carlo expectation is
observed in both neutral and doubly-charged combinations due to
imperfect simulation of the detector response.  The observed
data-to-simulation ratio from doubly-charged combinations is used to
correct the simulation of neutral pK combinations, shown in
Figure~\ref{fig:alephpk}(b).  The neutral pK combinations show clear
resonant activity in the mass range from 1460 to 1800\,MeV/$c^2$; this
is due to the production of known $\Lambda$ and $\Sigma^{*}$
resonances.  As illustrated in Figure~\ref{fig:alephpk}(c),
non-resonant simulation is subtracted from the data, and the resulting
invariant mass distribution is fit to the amplitude of eight NK
resonances, most prominently the $\Lambda(1520)$ followed by broader
contributions from $\Sigma(1480)$, $\Sigma(1560)$, $\Sigma(1580)$,
$\Sigma(1620)$, $\Sigma(1660)$, $\Sigma(1670)$, and $\Sigma(1750)$.
This results in a yield from the $\Lambda(1520)$ of
2874$\pm$320~(stat.)$\pm$270~(syst.)\,combinations.  When considered
with the selection efficiency of $9.7\pm0.9\%$ and
$\mathrm{Br}(\Lambda(1520)\rightarrow\mathrm{pK}^-)=$22.5\%, this
yields a production rate per hadronic Z decay of $N_{\Lambda(1520)} =
0.033\pm0.004\pm0.003$, in agreement with other measurements of this
quantity~\cite{lambdaOD}.  ALEPH then searched for resonant activity
in invariant mass distribution of pK$^{0}_{\mathrm{S}}$ combinations.
Figure~\ref{fig:alephpk0s} shows the distribution from 480\,000
combinations with a pK$^{0}_{\mathrm{S}}$ purity of 50\% compared to
the Monte Carlo simulation containing only known octet and decuplet
baryon ground states.  No evidence for the production of
$\Theta(1535)^+$ or its antiparticle is seen.  When considered with
the selection efficiency of $6.3\pm0.2\%$ and an assumed
$\mathrm{Br}(\Theta(1535)^+\rightarrow\mathrm{pK}^{0}_{\mathrm{S}})$=25\%
yields a 95\% C.L. upper limit on the production rate of
$\Theta(1535)^+$ and its antiparticle per hadronic Z decay is found to
be $N_{\Theta(1535)^+}<0.0025$.

\begin{figure}[tbp]
\includegraphics[width=7.5cm]{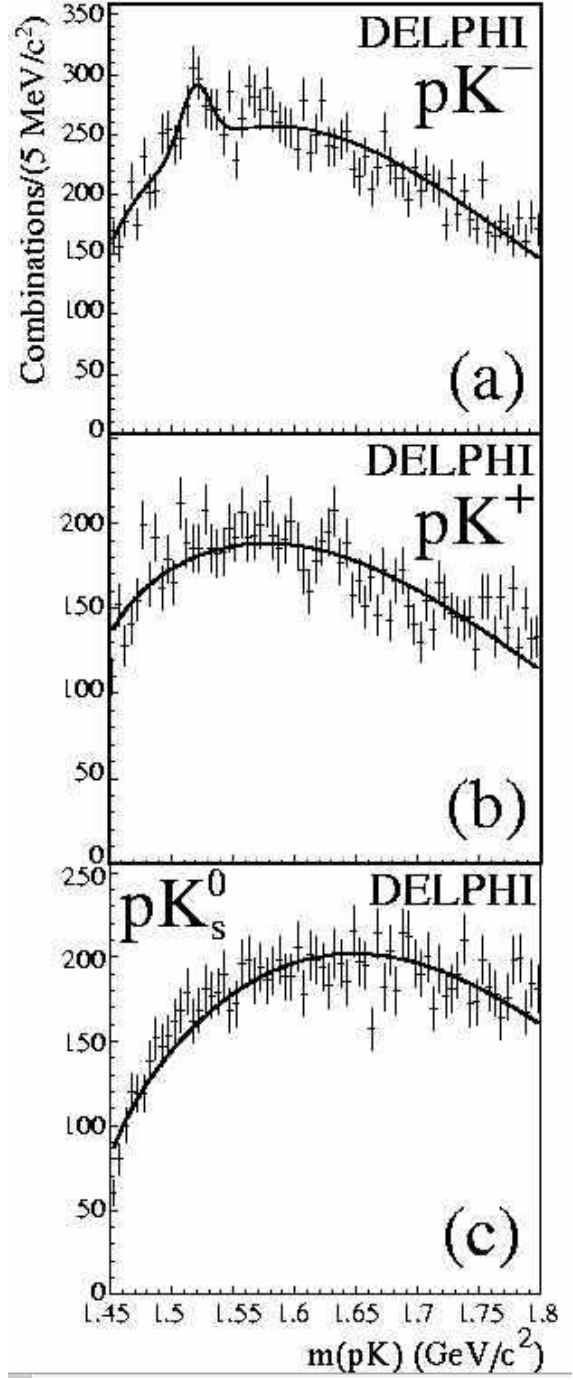}
\caption{DELPHI invariant mass spectra for (a) pK$^-$, (b) pK$^+$, and 
(c) pK$^0_{\mathrm{S}}$ combinations.  Data are denoted by crosses while a polynomial
fit is indicated by the solid line.  The peak in (a) corresponds to the
known $\Lambda(1520)$ resonance.}
\label{fig:delphipk}
\end{figure}

DELPHI have also examined invariant mass spectra of pK$^-$, pK$^+$,
and pK$^{0}_{\mathrm{S}}$ combinations from hadronic Z decays which
are shown in Figures~\ref{fig:delphipk}(a),~\ref{fig:delphipk}(b),
and~\ref{fig:delphipk}(c) respectively.  From a fit to the pK$^-$
spectrum, an excess in the $\Lambda(1520)$ region of
$306\pm55$\,events is observed consistent with the mass and width of
the $\Lambda(1520)$ as well as with previous measurements of the
production rate per hadronic Z decay~\cite{lambdaOD}.  No excess is
seen in the invariant mass spectrum for pK$^+$ combinations; this is
interpreted as an 95\% C.L. upper limit for the average multiplicity
of the production of a doubly-charged pentaquark, $\Theta^{++}$, of
$<N_{\Theta^{++}}>~<~0.006$ for the mass region between 1.5\,GeV/$c^2$
and 1.75\,GeV/$c^2$.  Furthermore, no excess is seen in the invariant
mass spectrum for pK$^{0}_{\mathrm{S}}$ and yields a 95\% C.L. upper
limit for the average multiplicity of $\Theta^+$ production of 
$<N_{\mathrm{\Theta^+}}>~<~0.015$ for the mass range between 1.52\,GeV/$c^2$ and 
1.56\,GeV/$c^2$

\begin{figure}[tbp]
\includegraphics[width=6cm]{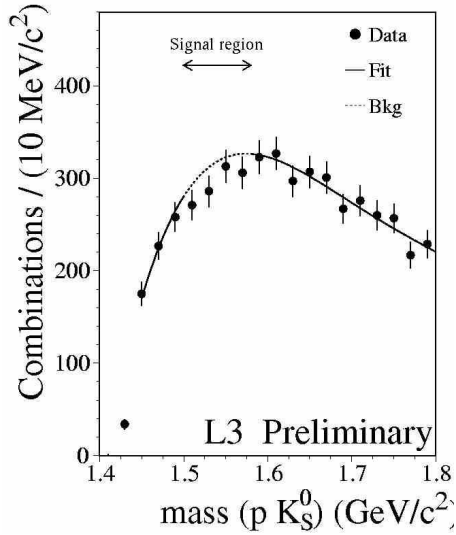}
\caption{L3 invariant mass spectrum for pK$^0_{\mathrm{S}}$ combinations.}
\label{fig:l3pk}
\end{figure}

L3 examined the invariant mass of pK$^0_{\mathrm{S}}$ combinations in
two-photon interactions; this invariant mass spectrum is shown in
Figure~\ref{fig:l3pk}.  No excess is observed in the data, with
1176\,events observed and $1253\pm44$\,events expected from a
background fit.  An 95\% C.L. upper limit for the inclusive production
reaction e$^+$e$^-\rightarrow\mathrm{e}^+\mathrm{e}^-\Theta^+X$ cross
section of about 20\,pb is obtained, to be compared to the inclusive
$\Lambda$ production cross section of $197.8\pm13.8$\,pb.

\subsection{Searches in $\Xi^{\pm}\pi^{\pm}$ channels}
\begin{figure}[tbp]
\includegraphics[width=7.5cm]{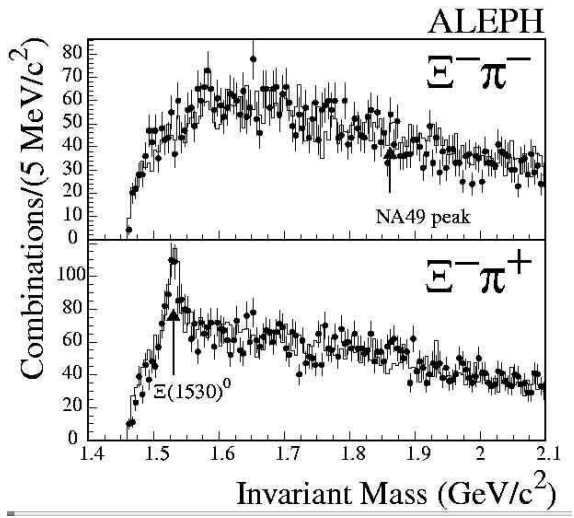}
\caption{ALEPH invariant mass spectra for (a)$\Xi^-\pi^-$ and (b)$\Xi^-\pi^+$ combinations
for the data (dots with error bars) and Monte Carlo simulation (histogram).}
\label{fig:alephcascade}
\end{figure}

ALEPH examined the invariant mass spectra of $\Xi^-\pi^-$ and
$\Xi^-\pi^+$ combinations; these distributions are shown in Figure~\ref{fig:alephcascade}.
In the $\Xi^-\pi^+$ distribution, a clear peak corresponding to the
production of the well-established $\Xi(1530)^0$ resonance is
observed and corresponds to a production rate per hadronic Z decay of
$N_{\Xi(1530)^0}=(77\pm8\pm6)\times10^{-4}$ in good agreement with
ALEPH and OPAL results~\cite{casoa}.  No evidence is seen for the
production of a doubly-charged or neutral $\Xi(1862)$ state, and
95\% C.L. upper limits are set at
\[
N_{\Xi(1862)^{--}}\mathrm{Br}(\Xi(1862)^{--}\rightarrow\Xi^-\pi^-) < 4.5\times10^{-4},
\]
\[
N_{\Xi(1862)^{0}}\mathrm{Br}(\Xi(1862)^{0}\rightarrow\Xi^-\pi^+) < 8.9\times10^{-4}.
\]

\subsection{Searches in D$^{\mathrm{(*)}\pm}$p channels}
\begin{figure}[tbp]
\includegraphics[width=7.5cm]{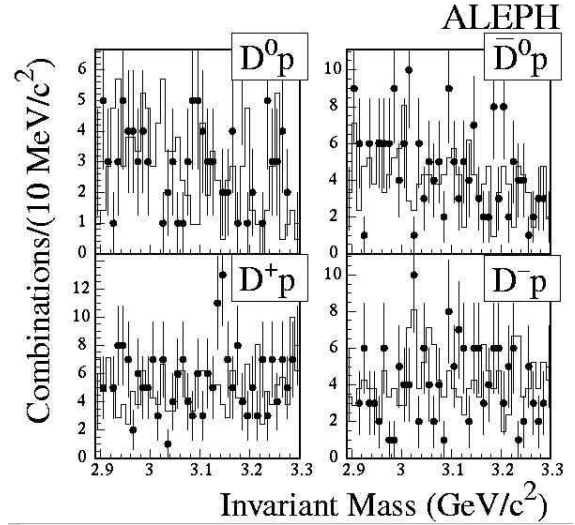}
\caption{ALEPH invariant mass spectra of Dp combinations for the data (dots with error bars)
and the Monte Carlo simulation (histogram).}
\label{fig:alephdp}
\end{figure}

ALEPH examined invariant mass spectra for Dp combinations (Figure~\ref{fig:alephdp}) and D$^{*\pm}$p
combinations for evidence of charm pentaquark $\Theta(3100)$ production.  No evidence was found.
\begin{figure}[tbp]
\includegraphics[width=7.5cm]{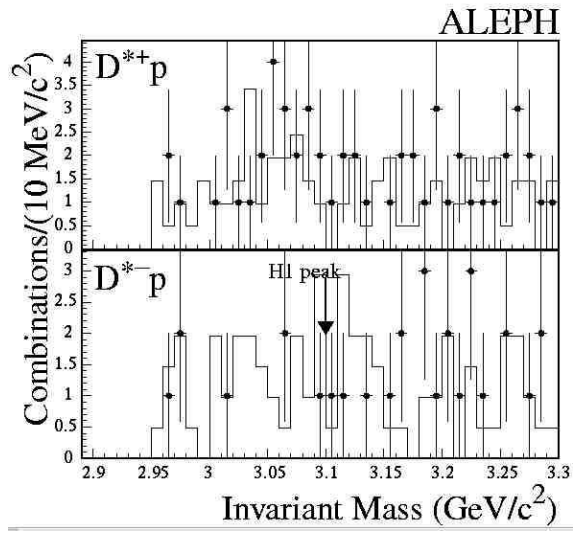}
\caption{ALEPH invariant mass spectra of D$^{*\pm}$p combinations for the data (dots with error
bars) and the Monte Carlo simulation (histogram).}
\label{fig:alephdstarp}
\end{figure}
Upper limits at 95\% C.L. were set on the production rates
\[
N_{\Theta_{c}(3100)^{0}}\mathrm{Br}(\Theta_{c}(3100)^{0}\rightarrow\mathrm{D}^{*-}\mathrm{p}) < 6.3\times10^{-4},
\]
\[
N_{\Theta_{c}(3100)^{0}}\mathrm{Br}(\Theta_{c}(3100)^{0}\rightarrow\mathrm{D}^{-}\mathrm{p}) < 31\times10^{-4}.
\]

\section{Summary}
No evidence for the production of pentaquarks in LEP data has been
observed by the ALEPH, DELPHI, and L3 experiments.  Production
scenarios involving fragmentation of quarks from hadronic Z decays as
well as inclusive production in two-photon interactions are
considered.  Upper limits at 95\% C.L. have been set on relevant
production rates.

\section{Acknowledgments}
This paper summarizes the work of and data collected by the ALEPH,
DELPHI, and L3 experimental collaborations.  We wish to thank our
colleagues from the CERN accelerator divisions for the successful
operation of LEP.  Furthermore, we are indebted to the engineers and
technicians from all the collaborating institutions for their support
in the construction and maintenance of the experiments.  The author
would like to thank CERN for its hospitality and acknowledge invaluable
discussions with and contributions from
Bertrand Echenard,
Peter Hansen,
Maria Kienzle,
Niels Jorgen Kjaer,
Mieczyslaw Krasny,
Chiara Mariotti,
Salvatore Mele,
Bruce Mellado, 
Efstathios Paganis,
Jack Steinberger,
Roberto Tenchini, 
Werner Wiedenmann, and
Sau Lan Wu.

\end{document}